\documentclass[twocolumn,showpacs,prl]{revtex4-1}
\usepackage{graphicx}
\usepackage{color}
\usepackage{amssymb}
\newcommand{\figline}[1]{(\begin{picture}(12,0)(0,0) 
\end{picture})}
\newcommand{\dotline}[1]{(\begin{picture}(12,0)(0,0) 
\end{picture})}
\begin{document}
\title{Light and heavy at the end of the funnel}
\author{Eef van Beveren$^{1}$ and George Rupp$^{2}$}
\affiliation{
$^{1}$Centro de F\'{\i}sica Computacional,
Departamento de F\'{\i}sica, Universidade de Coimbra,
P-3004-516 Coimbra, Portugal\\
$^{2}$Centro de F\'{\i}sica das Interac\c{c}\~{o}es Fundamentais,
Instituto Superior T\'{e}cnico, Edif\'{\i}cio Ci\^{e}ncia, Piso 3,
P-1049-001 Lisboa, Portugal
}
\date{\today}

\pacs{14.40.-n, 12.39.Pn, 11.80.Gw, 11.55.Ds}

\begin{abstract}
We show that, by taking a bare mass spectrum with constant spacings
for the quark-antiquark propagators, which is subject to considerable
mass shifts from meson loops, one adequately describes a large variety
of mesonic resonances, from the light scalars to the $b\bar{b}$ states.
All our results indicate that a harmonic-oscilator spectrum with
universal frequency, in combination with coupled-channel effects, does
a much better job than the $q\bar{q}$ spectrum of the funnel potential.
\end{abstract}

\maketitle

\section{Introduction}

For over three decades now, it has been widely assumed that the spectra
of quarkonia can be described by a kind of {\it universal},
{\it funnel-type} \/potential \cite{PREP200p127}, which basically stems
from the naive picture of color-flux-tube formation at large interquark
separations, while one-gluon exchange should represent the interactions
at shorter distances. The Cornell potential \cite{PRL34p369} was the
first and simplest version of such a potential.
However, several mesonic resonances observed long ago, as well as some
recently discovered ones, contradict such a description.
In particular, the $\rho$(1250--1300) (see e.g.\
Refs.~\cite{2ndSheetPoles,NPB76p375,NCA49p207,YF31p424,PZETF32p390,
PLB92p211,LNC37p236,YF41p1002,JINRE288521,NPPS21p105,SLACPUB5657,
SLACPUB5606,ARXIV07114748,NPA807p145})
does not at all fit in the $J^{PC}=1^{--}$ isovector spectrum of, for
instance, Ref.~\cite{PRD32p189}, which employed a semirelativistic version
of a funnel-type confining $q\bar{q}$ potential.
Moreover, evidence for the new vector charmonium
resonances $\psi(5S)(4790)$ and $\psi(4D)(4870)$
\cite{ARXIV10053490,PRD80p074001,EPL85p61002}, and
also for an $\Upsilon (4S)(10735)$ \cite{ARXIV09100967},
hints at masses that are again far from those
predicted by the confining funnel potential.
The same discrepancy between predictions of the funnel potential
and experiment is observed e.g.\ for the firmly established
$K^\ast(1410)$ \cite{PLB667p1}, as well as
the $f_2(1565)$ \cite{PLB667p1}.
Note that the discrepancies here are not of a few tens of MeV, but rather of
the order of 200~MeV!

Furthermore, the funnel potential does not accommodate the
light scalar mesons
$f_{0}(600)$, $K_{0}(800)$, $f_{0}(980)$, and $a_{0}(980)$,
nor the heavy-light scalar $D_{s0}(2317)$,
nor the $D_{s1}(2460)$ and the $D_{sJ}^*(2860)$,
at least not without unitarization \cite{ZPC30p615}.
But instead of unitarizing one's favorite quark model,
it has become fashion to fall back upon
tetraquarks, meson molecules, hybrids, or glueballs,
whenever a state does not fit the funnel potential
\cite{PRD79p114029,EPJC66p197}.

A better alternative to funnel confinement
has been suggested three decades ago \cite{PRD21p772}.
Based on the Weyl conformal invariance \cite{AdP59p101} property
of the theory of quark dynamics (QCD),
which leads, by a judicious choice of the time parameter,
to anti-De Sitter confinement (AdS) \cite{NCA80p401},
one obtains a harmonic-oscillator (HO) spectrum
for $q\bar{q}$ systems \cite{PRD30p1103}.
The experimentally observed spectrum is reproduced
via the inclusion of meson loops \cite{PRD21p772,PRD27p1527}.
Most remarkably, in the lowest-order approximation to AdS,
one obtains an interquark potential consisting of
a Coulomb-like term and a linear term \cite{LNP211p331},
exactly as in the case of lattice QCD (LQCD).
Nevertheless, full AdS yields an HO-like spectrum.
So it seems the lowest-order term of LQCD does not dictate
the spectrum of the $q\bar{q}$ propagator, as it does not provide
a satisfactory description of how confinement should follow from
full QCD.

\section{Harmonic oscillator and meson loops}

Unitarization of the quark model also reveals
threshold enhancements in electron-positron annihilation
\cite{AP324p1620}, which may explain e.g.\ why the $\psi (3770)$
consists of two strictures \cite{ARXIV08070494},
namely a $D\bar{D}$ threshold enhancement,
interfering with a $c\bar{c}$ resonance \cite{PRD80p074001}.
Furthermore, this phenomenon explains the enhancement
at 4.634 GeV observed by the Belle Collaboration in
$e^{+}e^{-}\to\Lambda_{c}^{+}\Lambda_{c}^{-}$ \cite{PRL101p172001},
not as a resonance, but rather a threshold enhancement
\cite{PRD80p074001}, or to be more precise,
at least three interfering threshold enhancements,
namely $\Lambda_{c}^{+}\Lambda_{c}^{-}$,
$D_{s1}(2536)^{\pm}D_{s}^{\ast\mp}$,
and $D_{s0}(2317)^{+}D_{s0}(2317)^{-}$.
Further threshold enhancements are observed in data published by
the BABAR Collaboration \cite{PRL95p142001}, though
not as peaked structures, but rather as valleys due to depletion
of the $e^{+}e^{-}\to J/\psi\pi^{+}\pi^{-}$ signal
at the opening of open-charm thresholds.
Moreover, at the positions
of $c\bar{c}$ resonances, one also observes dips in the data
where peaks should be expected \cite{PRL95p142001},
most noticeably at the $\psi (4S)$ resonance
(see Fig.~1 of Ref.~\cite{PRL105p102001}).
Assuming depletion as an explanation for the
$e^{+}e^{-}\to J/\psi\pi\pi$,
$e^{+}e^{-}\to J/\psi KK$, and $e^{+}e^{-}\to\psi (2S)\pi\pi$
signals, one is led to consider the corresponding new ``resonances''
$X(4260)$, $X(4660)$, \ldots leftovers
from open-charm decay \cite{PRL105p102001}.
Hence, the well-established $J^{PC}=1^{--}$ $c\bar{c}$ spectrum
anno 2010 still only consists of
$J/\psi$, $\psi (2S,3S,4S)$, and $\psi (1D,2D)$.

\begin{figure}[htbp]
\begin{tabular}{c}
\includegraphics[height=250pt]{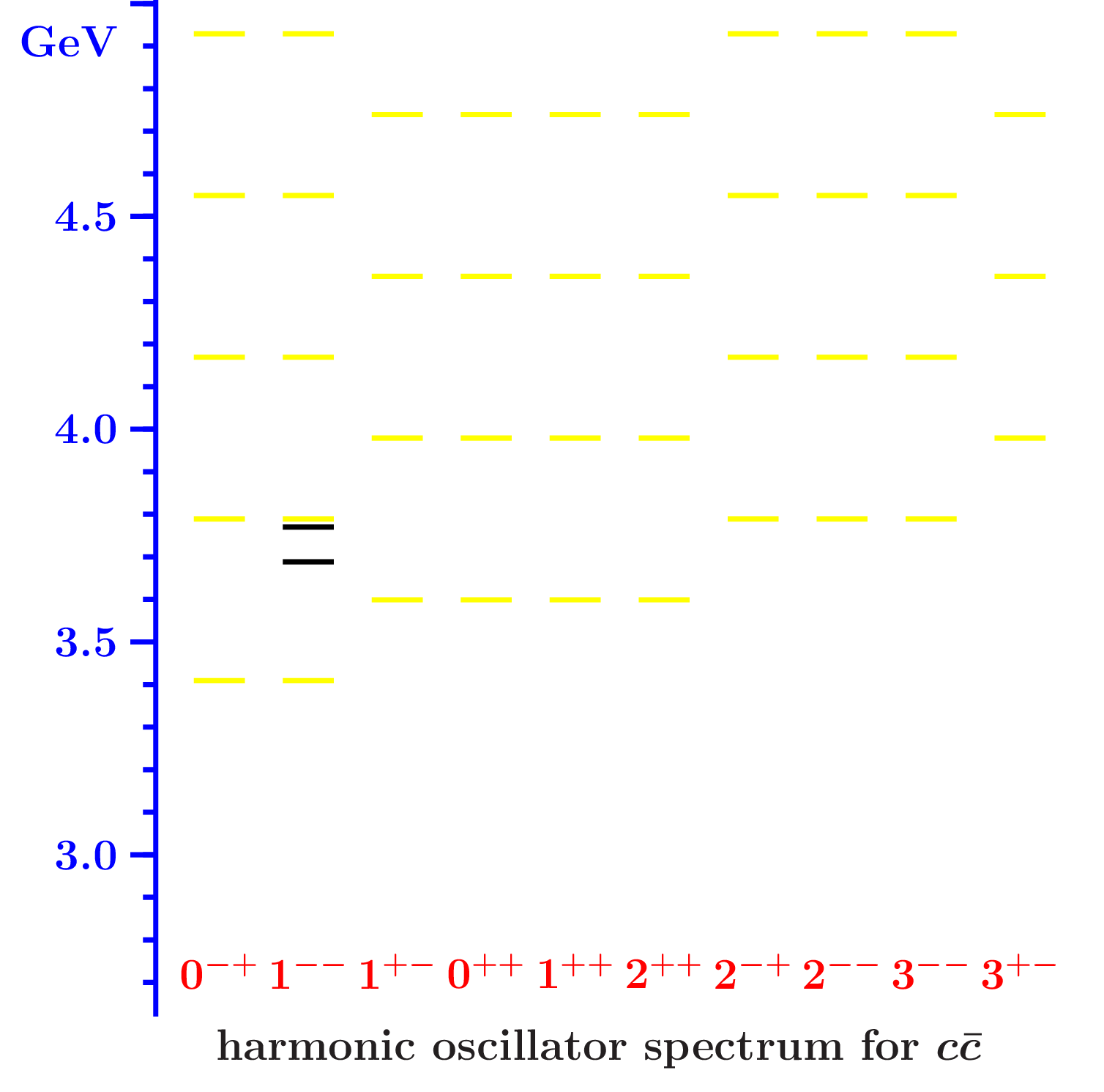}
\end{tabular}
\caption{
Charmonium level spectrum \figline{1 1 0}
as obtained from the HO, for
$m_{c}=1.562$, $\omega =0.190$ GeV.
The two black dashes indicate where the central masses for
the $\psi (2S)$ (lower dash) and the $\psi (1D)$ (upper dash)
end up through the effect of meson loops.
Similar mass shifts occur for other resonances and bound states.
}
\label{HO}
\end{figure}
Harmonic-oscillator confinement, with a radial level splitting
of 380 MeV, leads to abundantly many states of the $q\bar{q}$ propagator.
This is shown for $c\bar{c}$ in Fig.~\ref{HO}.
For other flavor combinations, the level scheme is exactly the same,
just shifted up/down according to
the sum of the effective quark masses.
In Fig.~\ref{HO}, we depict the effect
of meson loops on the central mass positions
of the $\psi (2S)$ and the $\psi (1D)$.
In the HO spectrum these two states are degenerate.
However, due to the interaction generated by the meson loops,
the poles associated with the resonances repel each other in such a way
that one of them is subject to a small mass shift,
whereas the other shifts considerably and downwards.
Higher up in the $c\bar{c}$ spectrum,
the mass shift of the lower pole becomes of the order of 150--200 MeV
\cite{PRD21p772}.
As a consequence, the associated resonance acquires a central mass
which is very similar to that of the $P$ states.
Observing and disentangling such states is not exactly an easy task,
as the past three decades have shown.

A way out is to study a well-isolated system, with just one
set of quantum numbers, like vector $c\bar{c}$,
which can be produced in $e^{+}e^{-}$ annihilation.
However, with a radial level separation of about 380 MeV
for the $S$ and $D$ states, combined with expected widths
of the order of 50-100 MeV, one needs a lot of good data,
binned in intervals of 5 MeV at most.
Furthermore, the opening of numerous thresholds complicates
such a task even more.
As a consequence, studying the lower part of the spectrum
is not the most adequate way to unravel the characteristics
of quark confinement,
since in particular the ground states suffer most from the
coupling to open charm, while sharp open-charm thresholds give
rise to noticeable enhancements.
What one really needs are the higher radial excitations,
very carefully distinguished from other effects,
like, e.g., threshold enhancements.
The latter occur less distinctly at higher energies, because they
involve open-charm mesons have larger widths, so that
thresholds are smeared out over several tens of MeV.
Finally, the phenomenon of threshold enhancements
also shows that specific open-charm channels decrease
rapidly in amplitude for higher invariant masses
\cite{PRD80p074001}.
Hence, studying the vector $c\bar{c}$ resonances in
open charm leads to
an extremely small set of data at higher energies.
Nevertheless, although the quality of $e^{+}e^{-}\to c\bar{c}$ data
is low for massive resonances, it is certainly justified to inspect
such data carefully, in order to find signs of further resonances.
Unfortunately, in practice lots of precious and, moreover, costly data
are simply discarded in many analyses,
by assuming an arbitrary background shape
(see {\it e.g.} Fig.~3 of Ref.~\cite{PRD80p091101}).
Three decades of very expensive accelerator physics
without any further results in the $c\bar{c}$ vector spectrum
is, to say the least, very embarrasing.

In Refs.~\cite{ARXIV10053490,PRL105p102001,ARXIV10044368,PRD80p074001,
ARXIV09062278,ARXIV09044351,EPL85p61002}
we have pointed at hints for the $\psi(3D)$, $\psi(5S)$,
$\psi(4D)$, $\psi(6S)$, and $\psi(5D)$ $c\bar{c}$ resonances in
several sets of reasonably good experimental data.
Furthermore, we have found the first indications \cite{ARXIV10053490} of
the $\psi(7S)$, $\psi(6D)$, and $\psi(8S)$ resonances
in recent data on $e^{+}e^{-}\to D^{\ast}\bar{D}^{\ast}$,
published by the BABAR Collaboration~\cite{PRD79p092001}.
\begin{figure}[htbp]
\begin{tabular}{c}
\includegraphics[height=250pt]{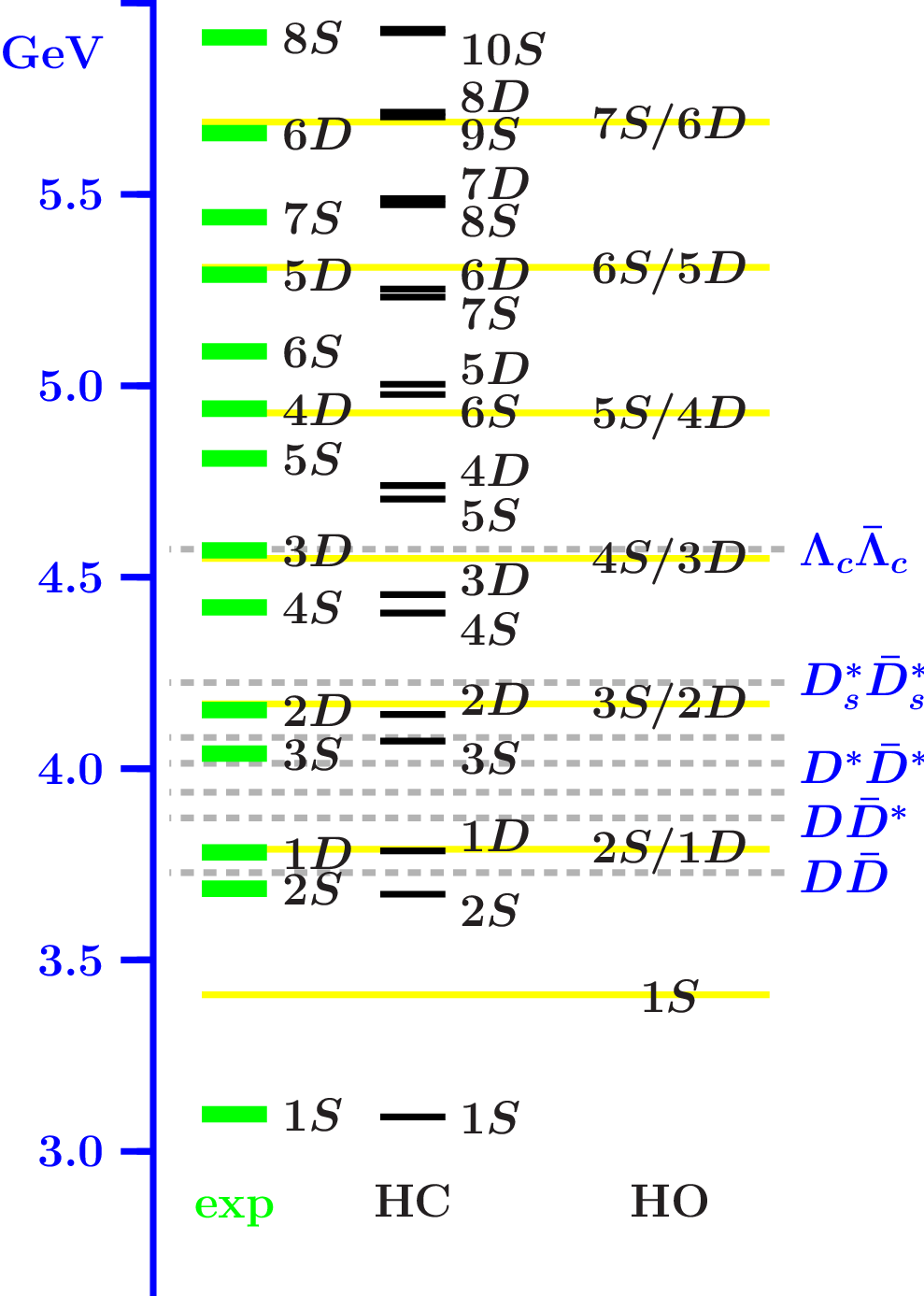}
\end{tabular}
\caption{
$J^{PC}=1^{--}$ $c\bar{c}$ level spectrum
observed by us in data from experiment (exp) \figline{0 1 0},
as predicted by the funnel-type $c\bar{c}$ potential model of
Ref.~\cite{PRD72p054026} (HC) \figline{0 0 0},
and as predicted by pure HO confinement
(HO) \figline{1 1 0}.
Meson and baryon loops shift the $D$ states a few MeV down/up,
whereas the $S$ states shift 100--200 MeV downwards.
For completeness, we also indicate the levels
of the sharp, low-lying meson-meson and baryon-baryon thresholds
\dotline{0.7 0.7 0.7} of the channels
$D\bar{D}$, $D\bar{D}^{\ast}$, $D_{s}\bar{D}_{s}$,
$D^{\ast}\bar{D}^{\ast}$, $D_{s}\bar{D}_{s}^{\ast}$,
$D_{s}^{\ast}\bar{D}_{s}^{\ast}$, and $\Lambda_{c}^{+}\Lambda_{c}^{-}$.
}
\label{HigherCharmonium}
\end{figure}
The resulting tentative spectrum of charmonium
is shown in Fig.~\ref{HigherCharmonium}.
It confirms the prediction of the model
of Ref.~\cite{PRD27p1527}, which is based on
HO confinement and meson loops, a model denoted
by us as HORSE.

In Fig.~\ref{HigherCharmonium} we have indicated the prediction of
the funnel-potential model of Ref.~\cite{PRD72p054026}.
The data of Fig.~\ref{HigherCharmonium} appear to contradict,
in particular,
the spin-orbit splittings as predicted in Ref.~\cite{PRD72p054026}.
In the latter model, the $S$-$D$ splittings for vector
$c\bar{c}$ states become smaller for higher radial excitations,
being only about 20 MeV for the $6D$-$7S$ mass difference. From
Fig.~\ref{HigherCharmonium} we estimate this splitting to be
roughly five to ten times larger.
Now, in the HORSE, $S$-$D$ mass differences are exactly zero
at the quenched level, but get generated by meson loops.
For the corresponding couplings,
the three-meson vertices determined in Ref.~\cite{ZPC21p291} are employed,
which involve the orbital and spin quantum numbers,
not only of the $c\bar{c}$ pair, but also of the mesons in the loops.
The resulting $S$-$D$ splittings come out very different then,
apart from the fact that the physical vector charmonium resonances naturally
appear as mixtures of $S$ and $D$ states.
We find that the combination dominated by the $D$ wave
shifts at most a few tens of MeVs away from the corresponding bare level.
The dominantly $S$-wave combination shifts substantially more,
viz.\ some 100--200 MeV, depending on the precise locations of nearby
thresholds.
This pattern is, to some extent, systematically repeated
for higher radial excitations, which the present data seem to confirm.

For the light-quark spectrum the situation is even more complicated,
as nonstrange and strange $q\bar{q}$ combinations
have comparable spectra, which will come out on top of each other, besides
possibly significant mixing of isoscalar $n\bar{n}$ ($n=u,d$) and $s\bar{s}$
states. To make things worse, decay channels involving kaons are common to
both $n\bar{n}$ and $s\bar{s}$ resonances.
The only system with a sufficient number of
established states to find evidence
(see Table 3 of Ref.~\cite{EPJA31p468})
for an HO level splitting of 380 MeV
is given by the radially excited $f_{2}$ mesons.
Moreover, the listing of experimental observations
and/or published analyses of the data
in the Particle Data Group (PDG) \cite{PLB667p1} tables
is not very helpful in the light-meson sector.
For example, whereas the HORSE clearly predicts the $\rho(2S)$
central mass at about 1.27 GeV,
the $\rho$(1250--1300) not even has its own entry. Observations of a
$\rho$-type resonance at 1.25--1.3 GeV \cite{PLB92p211}, which was moreover
confirmed in a recent multichannel analysis \cite{NPA807p145},
are listed under the $\rho$(1450), with the ``justification'':
\begin{quote}
Several observations on the $\omega\pi$ system in the 1200-MeV region
may be interpreted in terms of
either $J^{P}=1^{-}$ $\rho (770)\to\omega\pi$ production,
or $J^{P}=1^{+}$ $b_{1}$(1235) production.
We argue that no special entry for a $\rho$(1250) is needed.
The LASS amplitude analysis showing evidence for $\rho$(1270)
\cite{NPPS21p105} is preliminary and needs confirmation.
For completeness, the relevant observations
are listed under the $\rho$(1450).
\end{quote}
The resulting confusion is confirmed in Ref.~\cite{PRD82p036006},
where the authors, using dispersion relations, fail to observe
the $\rho (1270)$ resonance in data from
Refs.~\cite{PRD7p1279,NPB64p134,NPB79p301}.
Even more absurd is the lumping of $\phi(1500)$ and $\phi$(1900) observations
under the $\phi$(1680) \cite{PLB667p1}.

Finally, besides the resonance poles that follow directly from
$q\bar{q}$ confinement, meson loops may give also rise to dynamically
generated resonances, of which the low-lying scalar mesons
$f_{0}(600)$, $K_{0}(800)$, $f_{0}(980)$, and $a_{0}(980)$
were the first to be discovered \cite{ZPC30p615}.
But also the heavy-light scalar $D_{s0}(2317)$ is an example
of such a type of resonance.
Their existence will further complicate the classification
of experimentally observed resonances.
However, they can be studied and predicted by using the coupling of
HO confinement to the relevant meson-meson
and possibly baryon-baryon channels \cite{PRL91p012003}.

\section{Conclusion}

We have argued that the $c\bar{c}$ spectrum provides the best source to
infer the confinement mechanism, which then also allows
to understand and classify the spectra of other flavor
combinations. The available data clearly suggest that HO
confinement,  complemented with meson loops, is the best
candidate.

\section{Acknowledgments}

One of us (EvB) wishes to thank the organizers of {\it Menu 2010}
\/for their warm hospitality, and for bringing together many different
views on the strong interactions.
This work was supported by the {\it Funda\c{c}\~{a}o para a
Ci\^{e}ncia e a Tecnologia} \/of the {\it Minist\'{e}rio da Ci\^{e}ncia,
Tecnologia e Ensino Superior} \/of Portugal, under contract
CERN/FP/109307/2009.

\newcommand{\pubprt}[4]{#1 {\bf #2}, #3 (#4)}
\newcommand{\ertbid}[4]{[Erratum-ibid.~#1 {\bf #2}, #3 (#4)]}
\def\AdP{Annalen der Physik}
\def\AP{Ann.\ Phys.}
\def\EPJA{Eur.\ Phys.\ J.\ A}
\def\EPJC{Eur.\ Phys.\ J.\ C}
\def\EPL{Europhys.\ Lett.}
\def\LNC{Lett.\ Nuovo Cim.}
\def\LNP{Lect.\ Notes Phys.}
\def\NCA{Nuovo Cim.\ A}
\def\NPA{Nucl.\ Phys.\ A}
\def\NPB{Nucl.\ Phys.\ B}
\def\NPPS{Nucl.\ Phys.\ Proc.\ Suppl.}
\def\PLB{Phys.\ Lett.\ B}
\def\PRD{Phys.\ Rev.\ D}
\def\PRL{Phys.\ Rev.\ Lett.}
\def\PREP{Phys.\ Rept.}
\def\PZETF{Pisma Zh.\ Eksp.\ Teor.\ Fiz.}
\def\YF{Yad.\ Fiz.}
\def\ZPC{Z.\ Phys.\ C}

\end{document}